\documentclass[12pt,a4paper]{iopart}
\usepackage{amsfonts,iopams,cite}
\usepackage{mathrsfs}
\usepackage{graphicx}
\usepackage{dcolumn}  
\usepackage{psfrag}
\usepackage[usenames]{color}
\usepackage{hyperref}
\newcommand{\ket}[1]{\left\vert#1\right\rangle}
\newcommand{\bra}[1]{\left\langle#1\right\vert}

\newcommand{\braket}[2]{\langle{#1}|{#2}\rangle}

\newcommand{\beq}{\begin{equation}}
\newcommand{\eeq}{\end{equation}}

\renewcommand{\emph}[1]{{\it #1}}

\begin{document}
\title{Non-Gaussian distribution of collective operators in quantum spin chains}
\author{M. Moreno-Cardoner}
\address{Centre for Theoretical Atomic, Molecular and Optical Physics Queen's University, Belfast BT7 1NN, United Kingdom}
\author{J. F. Sherson}
\address{Department of Physics and Astronomy, Ny Munkegade 120, Aarhus University, 8000 Aarhus C, Denmark}
\author{G. De Chiara}
\address{Centre for Theoretical Atomic, Molecular and Optical Physics Queen's University, Belfast BT7 1NN, United Kingdom}
\date{\today}

\begin{abstract}
We numerically analyse the behavior of the full distribution of collective observables in quantum spin chains.
While most of previous studies of quantum critical phenomena are limited to the first moments, here we demonstrate how quantum fluctuations at criticality lead to highly non-Gaussian distributions.  Interestingly, we show that the distributions for different system sizes collapse on the same curve after scaling for a wide range of transitions: first and second order quantum transitions and transitions of the Berezinskii-Kosterlitz-Thouless type. We propose and analyse the feasibility of an experimental reconstruction of the distribution using light-matter interfaces for atoms in optical lattices or in optical resonators.
\end{abstract}
\maketitle

\section{Introduction}
The understanding of phase transitions and critical phenomena lies at the very heart of condensed-matter physics \cite{Sachdev}. Standard quantum phase transitions, i.e. those following Landau's theory, are signalled by the onset of a local order parameter when local symmetries are broken \cite{Landau}. This theory sets the mean value of the order parameter at the center of stage. However, going beyond the first order moment reveals interesting information about the many-body state without performing a full state tomography. For instance: in experiments with ultracold atoms, noise correlations reveal antiferromagnetic ordering \cite{Altman2004,Folling}; variances of collective operators in the form of structure factors allow to distinguish between quantum phases such as the superfluid and Mott insulator \cite{Rogers}, and to detect  antiferromagnetic or crystal ordering \cite{corco} and collective entanglement \cite{Meineke, Cramer}; the kurtosis, related to higher order moments, provides information about quasiparticle dynamics and their interactions \cite{Schweigler}; the full probability distribution of contrast in interference experiments can reveal strongly correlated atomic states \cite{Gritsev, Hofferberth}; and high-order correlation functions are needed to describe properly the physical outcome in the single-shots of several experiments of quantum many-body dynamics \cite{Sakmann}. Moreover, in recent experiments with trapped ions, the full-counting statistics of spin fluctuations allows for the detection of entangled, over-squeezed states \cite{Bohnet}.

In this paper we go beyond the first moments of the order parameter and analyse the full probability distribution function (PDF) of collective operators in spin chains. The  PDF of the order parameter plays a central role in statistical mechanics. When the correlation length is finite, deep in an ordered phase, 
the system can be regarded as the sum of independent subblocks of finite length and the central limit theorem leads to a Gaussian PDF, in agreement with Landau's paradigm.
Instead, at criticality, the divergence of the correlation length leads to a non-trivial highly non-Gaussian function, which characterizes the transition. If hyperscaling holds, the PDF shows finite size scaling with the universal critical exponents of the model  \cite{Cardy, Mussardo}.  Moreover, all statistical moments, related to many-body correlation functions, can be extracted from this function. The PDF thus contains non-local information of the system, and it can be connected to non-local order parameters in certain quantum phases. 

The PDF of magnetization has been exhaustively studied in classical spin systems undergoing phase transitions \cite{Binder,Bruce1,Bruce2,Bruce3,Hilfer,Tsypin,Antal,Archambault,Loison,Wysin}, and for those models, non-Gaussian distributions exhibiting finite size scaling at criticality have been found. However, much less attention has been paid to its quantum counterpart, and only few models such as the transverse Ising model have been previoulsy investigated \cite{Eisler,Lamacraft}. Non-Gaussian distributions of light fluctuations have been observed for cold atomic ensembles \cite{Mitchell}, but nevertheless, a systematic treatment in the case of strongly-correlated atoms is currently missing. In this context, quantum coherent fluctuations of the individual constituents can lead to non-Gaussian PDFs, which are a necessary resource for continuous variables quantum i nformation processing \cite{Weedbrook}.

The quantum-classical correspondence dictates for the quantum distribution the same finite size scaling with the corresponding critical exponents. However, to what extent the exact shape of the PDF  is universal or dependent on the microscopic details represents an intriguing question, which we address in this work. Here, using exact solutions and the density matrix renormalisation group (DMRG) \cite{DMRG,DMRG2}, we obtain the PDF of collective spin variables for different types of quantum phase transitions (first and second order, and BKT type), paying special attention to their behavior at criticality, and compare these results with their classical counterparts. Furthermore, we give an example on how the PDF can be connected to a non-local order parameter. 

In particular, we investigate realizations using ultra-cold atoms. Although at present relevant energy scales are still too small compared to the lowest temperatures achieved in current experimental setups, current efforts focused on finding efficient protocols to lower the total entropy of the system \cite{DeMarco}, for instance, by using Raman transitions \cite{ShersonCooling} or rehaping the trapping potential \cite{Kollath}, may well make such investigations feasible in the near future. Here, we propose two possible experimental setups based on optical lattices for the measurement of the PDF, the first employing high resolution microscopy and the second using quantum polarization spectroscopy.

\section{Probability density functions and finite size scaling} 
The probability to observe an eigenvalue $m$ of an operator $M$ is given by $P(m)=\sum_\mu \bra{m_\mu} \rho \ket{m_\mu}$, where $\rho$ is the density matrix describing the system and $\{\ket{m_\mu}\}$ is an orthonormal set of eigenstates of $M$ compatible with $m$. If $M$ is the order parameter, this distribution can be related  to the free energy functional $\mathcal{F}[m]$ appearing in Landau formalism \cite{Landau}: $P(m)\sim e^{-\mathcal{F}[m]}$. $\mathcal{F}[m]$ can be approximated by a power series of $m$ and, if the correlation length does not diverge, the lowest order terms dominate. To lowest order, this leads to a PDF which is approximately Gaussian, a result which can also be understood by the central limit theorem. In contrast, close to the critical point we expect a highly non-Gaussian PDF.

Close to a continuous (second order) phase transition induced by a parameter $g$ of the Hamiltonian with critical point $g_c$, the correlation length diverges as $\xi \propto \tau^{-\nu}$, where $\tau=|g-g_c|$ and $\nu$ is the critical exponent. Close enough to the critical point, the finite size scaling hypothesis implies that the mean value $\langle M\rangle$ scales with the system size $L$ as \cite{Fisher}:
\begin{eqnarray}
\langle M\rangle = L^{-\beta / \nu} f( \tau L^{1/\nu} )
\end{eqnarray}
where $f$ is an analytic function.
It is often more convenient and accurate for determining the critical point, both numerically and experimentally, to compute the Binder cumulant: 
\begin{equation}
U= 1-\frac{\langle M^4 \rangle}{3\langle M^2 \rangle^2},
\label{eq:Binder}
\end{equation}
as its scaling depends only on $\nu$ and not on $\beta$. That is, $U = \tilde f(\tau L^{1/\nu} )$, with a different scaling function $\tilde f$.  
$U$ quantifies the Gaussianity of the PDF, being null for a Gaussian distribution centered at zero. 

Here we go beyond the first moments and consider the full probability distribution function  $P_L(m,g)$. The renormalised PDF is expected to be a universal function (although still depending on the boundary conditions):
\begin{eqnarray}
\label{eq:rhoh}
\tilde P(\tilde m, r) =  L^{-\beta/\nu} P_L(m,g)
\end{eqnarray}
with $\tilde m=m / L^{-\beta/\nu}$ and $r=L/\xi$. Note that in this expression the parameter $g$ driving the second order phase transition needs to be changed when varying $L$, in order to keep $\xi/L$ fixed. This finite-size scaling of the PDF implies that hyperscaling and finite-size scaling of higher order correlation functions hold \cite{Binder,Bruce1,Tsypin}. In fact, and as we will see, when the critical exponents are unknown, or not defined in the standard way as in the BKT transition, one should instead rescale the PDF and the eigenvalue $m$  with $\sigma=\sqrt{\langle  M^2\rangle}$, directly computed from the PDF:
\begin{eqnarray}
\label{eq:rhoh}
\tilde P(\tilde m, g) =  \sigma P_L(m,g)
\end{eqnarray}
with $\tilde{m} = m/\sigma$.\\

\section{Models and Methods}
We consider a variety of spin-1/2 chain models encapsulated by the Hamiltonian:
\begin{eqnarray}
\label{eq:h}
H &=& \sum_{i,\alpha} \left(J_\alpha \sigma_\alpha^{i} \sigma_\alpha^{i+1} +h_\alpha \sigma_\alpha^i \right) 
\end{eqnarray}
where the sum on $i$ extends over the $L$ spins in the chain. The Pauli operators of spin $i$ are denoted by $\sigma_\alpha^i$, while $J_\alpha$ and $h_\alpha$ are coupling constants and magnetic fields along different directions ($\alpha=x,y,z$). This model exhibits various transitions \cite{Sachdev, Giamarchi, Mikeska}, some of which will be discussed below. We will study the behaviour of collective operators such as the total magnetization $M_\alpha=\sum_i \sigma_\alpha^i / L$ and its staggered counterpart $M^{st}_\alpha=\sum_i (-1)^i \sigma_\alpha^i / L$. 

Analytical results for the PDF can be obtained for models that can be written with a free fermionic representation after the Jordan-Wigner transformation, i.e. $J_z = h_{x,y} = 0$ in Eq.(\ref{eq:h}) assuming periodic boundary conditions, and operators that can be written as the sum of single-site operators in this basis, as for example $M_z$. 
The order  parameter $M_x$ however, is not separable and contains the string order operator. In principle all its powers could be computed by means of Wick's theorem, but the evaluation becomes involved as the order increases. At criticality and for the quantum Ising spin chain, the PDF can be exactly obtained by exploiting the relation with Kondo physics \cite{Lamacraft}, but in general, reconstructing the PDF analytically is a very challenging task. Here instead, to obtain the PDF, we combine two different numerical methods. We use exact diagonalisation for sizes up to 20 sites, and also the time-dependent density matrix renormalisation group (t-DMRG) \cite{DMRG} with open boundary conditions. The t-DMRG method provides an efficient way to evaluate the characteristic function, defined as the Fourier transform of the PDF. For a pure quantum state $\ket{\Psi_0}$ this can be written as:
\beq 
\chi(u) =  \sum_{m} P(m) e^{i u m} =  \sum_m |\braket{m}{\Psi_0}|^2 e^{i u m} =\bra{\psi_0} e^{i u M} \ket{\psi_0}
\eeq
being $M$ the operator of interest and $\left\{ m \right\}$ the corresponding set of eigenvalues. The expression for $\chi(u)$ is equivalent to the overlap between the initial state $\ket{\psi_0}$ and its evolution under a fictitious Hamiltonian equal to $M$ at different  times $u$, and thus, it can be computed with t-DMRG. Since there is a one-to-one correspondence between the characteristic function and the PDF, this can be recovered by inverse Fourier transforming $\chi(u)$. Moreover, it allows to directly evaluate all central momenta and cumulants.\\

\section{Results}
\subsection{Second order quantum phase transition}
\begin{figure}[h!]
\centering
\includegraphics[width=12cm]{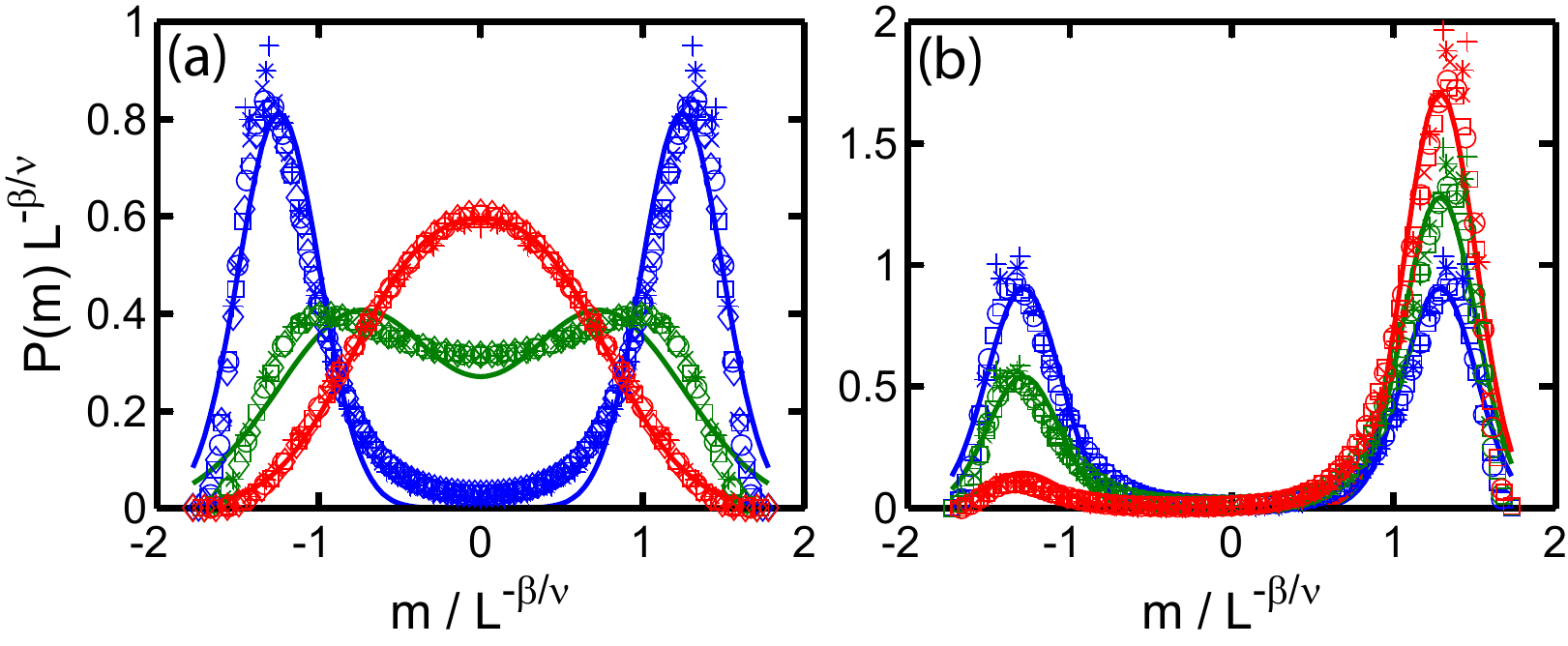}
\caption{(Color online) Rescaled PDF of the order parameter in the transverse Ising model. Data collapse is observed for different system sizes ($L=20,30,40,60,80,100$ correspond to +, star, x, circle, square and diamond symbol, respectively). (a) In absence of longitudinal field and for different phases at fixed $r$: FM (blue) at $r=4$ (equivalent to $g=0.75$ for $L=16$), critical (green) at $r=0$ ($g=1.0$) and PM (red) (at $r=4$, equivalent to $g=1.25$ for $L=16$). (b) In presence of the longitudinal field in the FM phase at $r=4.8$ (equivalent to $g=0.7$ for $L=16$) and $q=0$ ($h=0$, blue), $q=0.5$ (green, equivalent to $h=6\cdot 10^{-5}$ for $L=16$) and $q=2$ (red, equivalent to $h=2\cdot 10^{-4}$ for $L=16$). The solid lines correspond to a Gaussian fit as discussed in the text. The PDF is non-Gaussian close to the critical point.}
\label{Figure1}
\end{figure}

We set $J_{y,z}=h_{x,y}=0$, $J_x=J<0$ and $h_z=Jg$ in Eq.(\ref{eq:h}), corresponding to the ferromagnetic transverse Ising model, which exhibits a second order phase transition at $g=\pm 1$  separating a ferromagnetic ordered phase (FM) at low fields and a paramagnetic disordered phase (PM) at high fields.
In Fig. \ref{Figure1}(a) we show the numerical results for the PDF of the spontaneous magnetization $M_x$, the order parameter in FM. Remarkably, we have obtained data collapse already for  relatively small system sizes, assuming the ansatz of Eq.(\ref{eq:rhoh}) with the predicted values $\beta=1/8$  and $\nu=1$ (see e.g. \cite{Cardy, Mussardo}). This result reinforces the scaling hypothesis of all the statistical moments of the order parameter. Away from the critical point and in the thermodynamic limit ($r\rightarrow  \infty$) the PDF is Gaussian, in agreement with Landau theory and the central limit theorem. In the disordered phase, it is centered around $m=0$, whereas in the ordered phase the distribution is bimodal with the two peaks corresponding approximately to the broken-symmetry values of the order parameter. In contrast, at criticality ($r\rightarrow 0$) the distribution becomes non-Gaussian. The central limit theorem does not necessarily hold anymore due to quasi long-range fluctuations in this regime. We note that, despite the PDF is universal at criticality, it still depends on the boundary conditions that are chosen. 

For comparison, it is worth noting that for the transverse magnetization $M_z$, which is not the order parameter, the PDF is always Gaussian. For this observable, the moment of $n$-th order can be directly evaluated from the characteristic function and it always scales as $L^{1-n}$. This leads to a Gaussian distribution in the thermodynamic limit.\\

As previously discussed, it is not clear a priori that the PDF is a universal function (only depending on the universality class), and thus, that it can be retrieved from the classical model analogue. The 1D transverse Ising model maps onto the classical 2D Ising model with spatially anisotropic couplings. The anisotropy depends on the temperature of the quantum system, and therefore, on the system size of the equivalent classical model. Performing finite size scaling in the 2D model simultaneously in all dimensions (as it is usually done for a $z=1$ transition) changes the anisotropy at each step, and one might doubt whether this process leads  to a universal function. To elucidate this, we compare the previous results with those obtained for the classical model by Monte Carlo in \cite{Hilfer,MARTINS2004}. A direct comparison shows that the quantum results for the PDF at the critical point seem to have more fluctuations around $m=0$, leading to a flatter distribution with non-zero value, in contrast to the classical case.

\begin{figure}[h!]
\centering
\includegraphics[width=15cm]{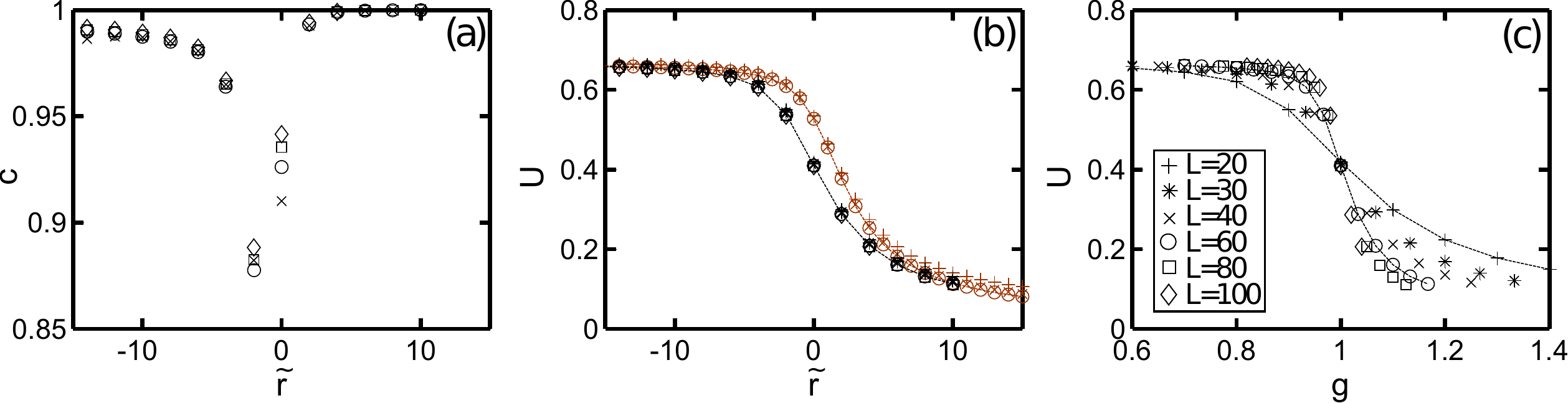}
\caption{(Color Online) (a) Correlation coefficient $c$ of a non-linear least square fitting of the PDF $P(m)$ to the Gaussian function $P_\textrm{fit}(m)$ in the transverse Ising model. The coefficient $c$ suddenly decreases close to the critical point, indicating the PDF is non-Gaussian. (b-c) Binder Cumulant $U$ in the Transverse Ising model, for different system sizes denoted by different symbols (from $L=20$ to $L=100$). (b) As a function of $\tilde{r}=L/\xi\cdot \textrm{sgn}(g-g_c)=L(g-g_c)$, data collapse is observed. (c) As a function of $g$ for OBC, the lines cross at the critical point. Black and brown symbols are for OBC and PBC respectively.}
\label{Figure2}
\end{figure}

\textit{\textbf{Gaussianity of the PDF.--}} We can characterize the gaussianity of the distribution by fitting the data with the sum of two Gaussians as 
\begin{equation}
P_\textrm{fit}(m)=\frac{1}{2\sqrt{2\pi \sigma^2}} 
\left[e^{-(m-m_{0})^2 / (2\sigma^2)}+ e^{-(m+m_{0})^2 / (2\sigma^2)}\right]
\end{equation}
with $\sigma^2$ and $m_0$ as free parameters (see solid lines in Fig.~\ref{Figure1}(a)). As shown in the figure, we find very good agreement away from the critical point, whereas close to the critical point the fitting yields poor results and one should use instead the exponential of a higher-order polynomial. This can be quantified by the non-linear least square fitting correlation coefficient defined as:
\begin{equation}
c=\sqrt{1-\delta y^2 / \textrm{Var}[y]}
\end{equation}
where $\delta y^2$ is the squared norm of the residuals of the data $y$ and $\textrm{Var}[y] $ the corresponding variance. This quantity is close to one when the fitting works well, whereas it drops to smaller values for a poor fitting. Fig.~\ref{Figure2}(a) shows $c$ as a function of $\tilde{r}=r~ \textrm{sgn}(g-1)= L(g-1)$. Clearly, there is a sudden decrease close to the critical point ($\tilde{r}=0$), whereas it tends to one away from the critical point ($\tilde{r}\rightarrow \pm \infty$).

An alternative way to quantify the Gaussianity of the PDF is by evaluating the Binder cumulant $U$, Eq.(\ref{eq:Binder}). $U$ vanishes for a Gaussian distribution centered at zero, whereas it tends to $U\rightarrow 2/3$ for a distribution close to two symmetric delta functions. Evaluating $U$ is convenient for locating the critical point, since its finite size scaling only depends on the critical exponent $\nu$ and not on $\beta$, that is, $U=\tilde{f}(r)$,  and in this case $U=\tilde{f} (|g-g_c| \cdot L)$. Thus, when plotted as a function of the transverse field $g$, the crossing point for data corresponding to different sizes tends to the actual critical point $g_c$. When instead, plotted as a function of $\tilde{r}$, they collapse to the same curve,  which depends, however, on the boundary conditions. 

In the same figure we also plot the result of $U$ for the transverse Ising model with open boundary conditions (OBC) as a function of $\tilde{r}$ (b) and $g$ (c) for different system sizes, ranging from $L=20$ to $L=100$. The result for periodic boundary conditions (PBC) is also shown for comparison.\\
 
\subsection{First order phase transition}
We consider the same Hamiltonian as before but with an additional longitudinal field $h_x=J h \neq 0$. At fixed value of the transverse field $g=\pm 1$ and approaching the Ising critical point by  varying only $h$, the quantum phase transition is characterized by a different set of exponents ($\beta=1/15$ and $\nu=8/15$ \cite{Cardy}). Our results (not shown) support strong evidence of scaling of the PDF. 

If instead, the longitudinal field $h$ is varied across zero, but at fixed value of the transverse field $|g|<1$, the system undergoes a first order transition between two ferromagnetic states with opposite magnetization $M_x$.  At this transition, neither the correlation length diverges nor the gap closes. Nevertheless, inspired by Ref.~\cite{Campostrini2014}, we propose a finite size scaling of the PDF.
 
In a finite size chain, two different energy gaps are present in this model. One is the real energy gap in the thermodynamic limit, which at $h=0$ is given by $\Delta E = | |g|-1|^{\nu z}$ with $z=1$, and it closes at the Ising critical point ($g=\pm 1$). The other gap $\delta$ separates the two lowest eigenstates and it is minimum at $h=0$, with value $\delta_0 = 2|J| (1-g^2) g^L$, decreasing exponentially with $L$ ($|g|<1$). In \cite{Campostrini2014}, following dimensional arguments, the authors assume that $\delta$ and $M_x$ only depend on the longitudinal field $h$ and $L$ through the ratio $q$ between the two energy scales: the first associated with the longitudinal field $h L M_{x,0}$, being $M_{x,0} = (1-g^2)^{1/8}$ the magnetization at $h=0$, and the second $\delta_0$:
\begin{equation}
q = \frac{2 h L M_{x,0}}{\delta_0}= \frac{h L}{(1-g^2)^{7/8} g^L}.
\label{Eq:kappa}
\end{equation}
Thus, $\delta \sim \delta_0(g) f_1(q)$ and $M_{x} \sim m_0(g) f_2(q)$, where $f_{1,2}(q)$ are analytic functions at $q=0$.

Here, we conjecture a similar dependence for the correlation length:
$\xi \sim \xi_0(g)  f(q) $ where $f$ is analytic at $q=0$, and $\xi_0(g)=\xi(q=0,g)$ diverges close to the Ising critical point with the usual power law: $\xi_0(g) \sim \left| |g| - 1 \right|^{-\nu}$. 
Moreover, any observable $M_L(g,h)$  depends only on the ratios $r=L/\xi_0$ and $q$ when rescaled with the proper critical exponents, e.g. for the magnetization $\tilde{M}_x(r,q)=L^{-\beta/\nu} M_{x,L}(g,h)$. Indeed, at fixed values of $r$ and $q$ (and for large enough chains, or equivalently weak enough external fields), we observe again data collapse for the PDF for different lengths, as shown in Fig. \ref{Figure1}(b). 
We find the distribution to be always bimodal with the relative height of the two peaks ruled by $h$. We fitted the data with the sum of two Gaussians  and found reasonable agreement except at the critical point, due to non-linear effects in the Landau potential.

\subsection{BKT transition}
\begin{figure}[h!]
\centering
\includegraphics[width=12cm]{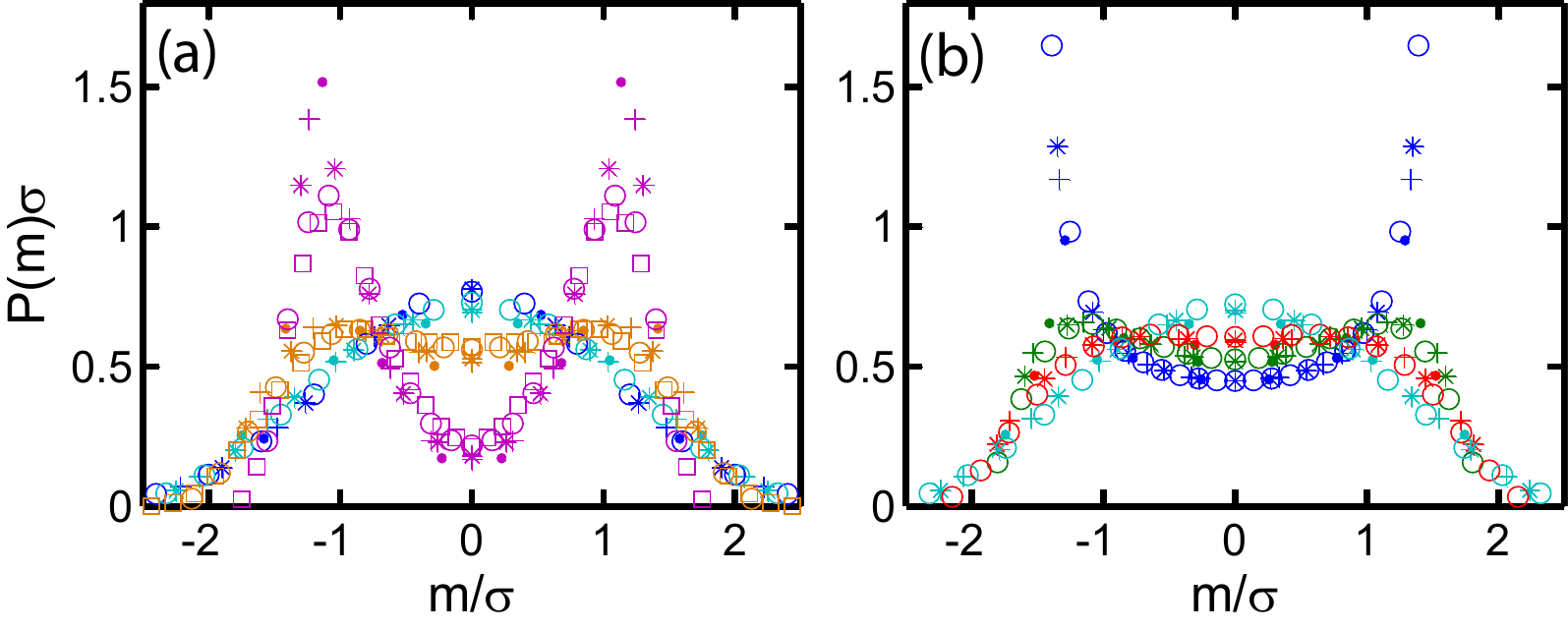}
\caption{(Color online) Rescaled PDF in the spin-$1/2$ XXZ model. 
Data collapse is observed for different system sizes ($L=10,20,30,40,60$ correspond to dot, cross, star, circle and square symbol, respectively). (a) For $M_z^{st}$ at fixed $r$: (critical phase)  $r=0.9$ (blue) and $r=0$ (cyan); (N\'eel ordered phase) $r=0.35$ (orange) and $r=1.3$ (purple). (b) For $M_x^{st}$ at fixed $\Delta$: (critical phase) $\Delta=-0.99$ (blue), $\Delta=-0.5$ (green), $\Delta=0$ (red) and $\Delta=1$ (cyan).}
\label{Figure3}
\end{figure}

For completeness, we finish this analysis with a different type of transition, the Berezinski-Kosterlitz-Thouless (BKT) transition in the spin-$1/2$ XXZ model. We set $h_{x,y,z}=0$, $J_{x,y}=J>0$ and $J_z=J\Delta $ in Hamiltonian (\ref{eq:h}). The phase diagram is well known: the ground state is ferromagnetic for $\Delta<-1$, critical for $|\Delta|<1$, and with N\'eel order for $\Delta>1$. The BKT transition is at $\Delta=1$. This transition is of the same universality class as the classical 2D XY model, for which previous calculations on the PDF \cite{Archambault} and on the Binder cumulant \cite{Loison,Wysin}, show that the order parameter distribution is non-Gaussian at criticality. Here we compute the PDF for the ground state of the XXZ model, for both staggered magnetizations $M_{x}^{st}$ and $M_{z}^{st}$,  corresponding to the order parameters in the critical and N\'eel phases respectively, and for the full range of values of $\Delta$. In contrast to the previous sections, we rescale the quantities $m$ and $P(m)$ with $\sigma$ as defined in Eq.(\ref{eq:rhoh}). For $M^{st}_z$ we also fix $r=L/\xi$, being $\xi = e^{\pi/\sqrt{|\Delta -1|}}$ \cite{Mikeska}, and observe data collapse for different system sizes, as shown in Fig.~\ref{Figure3}(a). As expected, the distribution tends in the thermodynamic limit to a double- and single-peaked Gaussian for $\Delta>1$ (ordered phase with respect to $M_z^{st}$) and $\Delta<1$, respectively. 

As shown in Fig.~\ref{Figure3}(b), the situation is much more interesting for the PDF of $M^{st}_x$, for which the spin-spin correlations do not decay exponentially in the critical phase ($|\Delta|<1$). Outside this interval the distribution is Gaussian and centered at zero, because the operator  $M^{st}_x$ is disordered, while in the critical phase it is highly non-Gaussian. We observe again data collapse for different chain lengths, but now when fixing the value of $\Delta$. The scaling with fixed $\Delta$ is expected to occur in a critical phase, where the energy gap has already closed. When crossing the first order transition at $\Delta = -1$, the PDF shows also a discontinuity, with a sudden jump from a Gaussian to a function that has a singular behavior. In contrast to the usual behavior of the PDF, this function does not tend to zero as $m/\sigma \rightarrow \pm \infty$, but it increases its value and its derivative when increasing the system size (see blue symbols in Figure \ref{Figure3}(b)). Across the critical phase, the function changes continuously from a double-peak structure for $\Delta<0$ to a single peak distribution for  $\Delta>0$. At the BKT point the PDF tends again to a Gaussian in the thermodynamic limit. Note that the results for the ferromagnetic model $J<0$ are equivalent when analyzing $M_{x}$ instead, and changing $\Delta \leftrightarrow -\Delta$. 

We emphasise that while the collapse is expected with the correct universal exponents as in the classical case, the specific collapse function can be dependent on the model or on the  boundary conditions. This is clear by visual inspection when comparing the previous results at different values of $\Delta$ to those in the 2D classical XY model reported in \cite{Archambault} (Figure 2).

\begin{figure}[h!]
\centering
\includegraphics[width=12cm]{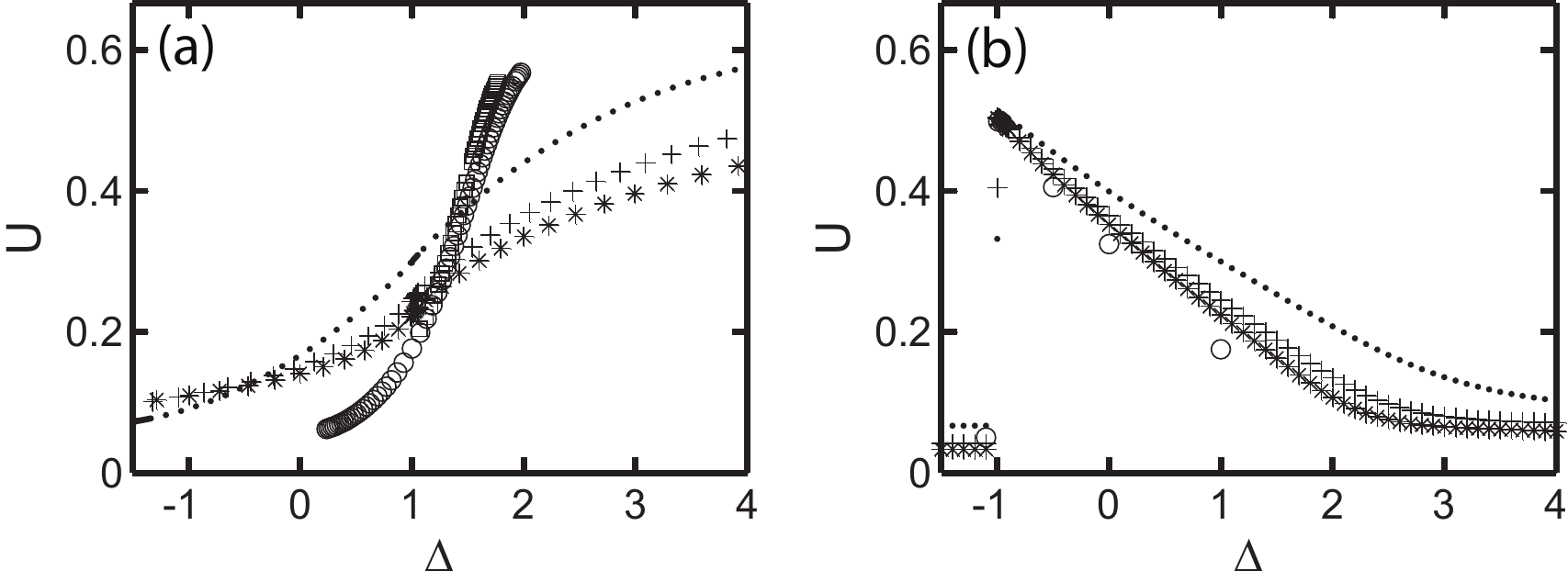}
\caption{Binder Cumulant $U$ in the spin-$1/2$ XXZ model, for different system sizes denoted by different symbols ($L=10,20,30,40,60$ correspond to dot, cross, star, circle and square symbol, respectively). (a) For $M_z^{st}$ it tends to $U=2/3$ in the N\'eel ordered phase in the thermodynamic limit. (b) For $M_x^{st}$ is finite in the critical phase in the thermodynamic limit.}
\label{Figure4}
\end{figure}

The non-Gaussianity in the critical phase is also evident from the analysis of the Binder cumulant. In Fig.~\ref{Figure4} we show the result of $U$ for the spin-$1/2$ XXZ model for the two observables $M_z^{st}$ (left) and $M_x^{st}$ (right), as a function of $\Delta$. The Binder cumulant for $M_z^{st}$ tends to the value $U\rightarrow 2/3$ in the N\'eel phase in the thermodynamical limit, whereas the one for $M_x^{st}$ remains finite in the critical phase. 

\section{Non-local order parameters}
The BKT transition is particularly relevant in the context of 1D optical lattice gases because it rules the superfluid (SF) to Mott insulator (MI) transition in the Bose-Hubbard model. Close to the MI, where density fluctuations are small, and for large enough integer fillings, the model can be approximated to an effective spin-$1$ model \cite{Altman1, Huber}. The SF and MI are mapped respectively to the critical ferromagnetic phase, and to a state with local magnetization $S_z^i = 0$ perturbed with tightly bound particle-hole fluctuations. The nature of density fluctuations is however different in the two  phases. In the MI the non-local correlations can be characterized by the parity operator defined as 
\begin{equation}
\mathcal{O}^2_P = \lim_{l\rightarrow \infty} \left\langle \prod_{k\leq j \leq k+l} e^{i\pi\delta n_j} \right\rangle
\end{equation}
where $\delta n_j$ is the local excess density from the average filling \cite{DallaTorre1,DallaTorre2,Endres}, and corresponds to the local magnetization in the magnetic model. $\mathcal{O}^2_P$ is finite in the MI, while it vanishes in the SF, and it has been experimentally reconstructed in \cite{Endres} using single site microscopy.

In this context, an additional motivation for the study of PDFs is that it captures these non-local correlations. Indeed, it is easy to see that the characteristic function 
$X(u) = \sum_m e^{i u m} P(m)$
for the suitable collective variable (in this case, the average magnetization of a subblock of length $l$) at frequency $u=l \pi$ is equal to the parity operator $\lim_{l\rightarrow \infty} X(u=l\pi) = \mathcal{O}^2_P$. Thus, a measurement of a collective variable and the reconstruction of its PDF provides an alternative route to evaluate some particular non-local order parameters. 

\section{Experimental reconstruction of the PDF} 
The PDF for the local number parity operator can be reconstructed using single site resolution microscopy \cite{Greiner2009,ShersonKuhr}. Recently, novel schemes which allow to circumvent light-assisted pair loss and resolve the internal atomic degree of freedom have been also demonstrated \cite{Greiner2015}. An alternative proposal is based on the Faraday effect, and consists in analysing the polarization fluctuations of a strongly polarised laser beam that interacts with the atomic sample \cite{Polzik1,Roscilde,Eckert}. Since this is based on dispersive light-matter interaction, it has the advantage of being less destructive, which could potentially pave the way for multiple probing of the same ensemble. Thus, this method could find relevant applications for quantum state control and preparation. In particular, by monitoring the system in the appropriate way, this could be used in principle to lower its total entropy. Already it has been experimentally demonstrated that QPS combined with real-time feedback allows for a dramatic reduction of  the atom number fluctuations \cite{ShersonFeedback}, increasing the purtity of a coherent spin state \cite{Bouyer}, and inducing spin squeezing \cite{Mitchell}. Moreover, QPS can be exploited for quantum state engineering and control \cite{ShersonControl}, inducing non-trivial quantum dynamics \cite{Meshkov}, and the conversion of atomic correlations and entanglement into the light degree of freedom for quantum information processing \cite{ShersonMolmer}.

The light-matter interaction with the collective spin operator $M$, $H_\textrm{eff} \sim (\tilde{\kappa}/\tau) L P_{ph} M$, 
is written in terms of the momentum-like light quadrature $P_{ph}$ measuring the photon fluctuations in the circular basis with respect to the strong polarisation axis, and $\tau$ is the interaction time. $P_{ph}$ is canonically conjugated to its position-like counterpart: $[X_{ph},P_{ph}]=i\hbar$. The coupling constant $\kappa = (\eta \alpha)^{1/2}$ can be expressed in terms of the single atom excitation probability or destructivity $\eta$, and the resonant optical depth $\alpha=N \sigma_{cross}/A$, where $A$ is the overlap area between light and atoms, $\sigma_{cross}$ is the resonant cross section and for a one-dimensional system $N=L/d$, being $d$ the interparticle distance. By adjusting the laser intensity, the destructivity parameter $\eta$ is typically set to values smaller than $0.1$ to limit the fraction of excited atoms. The resonant optical depth $\alpha$ should be maximized in order to obtain the largest possible coupling between light and atoms. For ultracold atoms in a one-dimensional optical lattice, considering $L=100$ and $A=0.5 \mu m^2$ we obtain $\alpha\approx 8$ and $\kappa\approx 1$. We define $\tilde\kappa=\kappa / \sqrt{L}$, which will be approximately independent of $L$ and could be (in the best case scenario) as large as $\tilde{\kappa} \sim 0.1$. We will show that this value of $\kappa$ is large enough to reconstruct the PDF of the spin operator $M$ by looking at the distribution of the light quadrature $X_{ph}$. Larger values of $\tilde{\kappa}$ could be engineered by coupling atoms with optical cavities \cite{Vuletic1,Vuletic2}, nanophotonic crystals \cite{Chang1,Chang2} or optical nanofibers \cite{Rauschen}.

It is possible to show \cite{DeChiara2015} that the light distribution is the sum of vacuum Gaussian distributions of light each displaced by a quantity proportional to the eigenvalue $m$ of the operator $M$ and scaled by the probability $P(m)$ to observe such eigenvalue:
\begin{eqnarray}
P(X_{ph})=\sqrt{\frac{1}{2\pi \sigma_{ph}^2}}\sum_{m} P(m) e^{-(X_{ph}+\tilde{\kappa} N m)^2 / (2\sigma_{ph}^2)}
\end{eqnarray}
where we have considered a Gaussian input light beam with variance $\sigma_{ph}^2$, being $1/2$ for the vacuum state.  

\begin{figure}[h!]
\centering
\includegraphics[width=12cm]{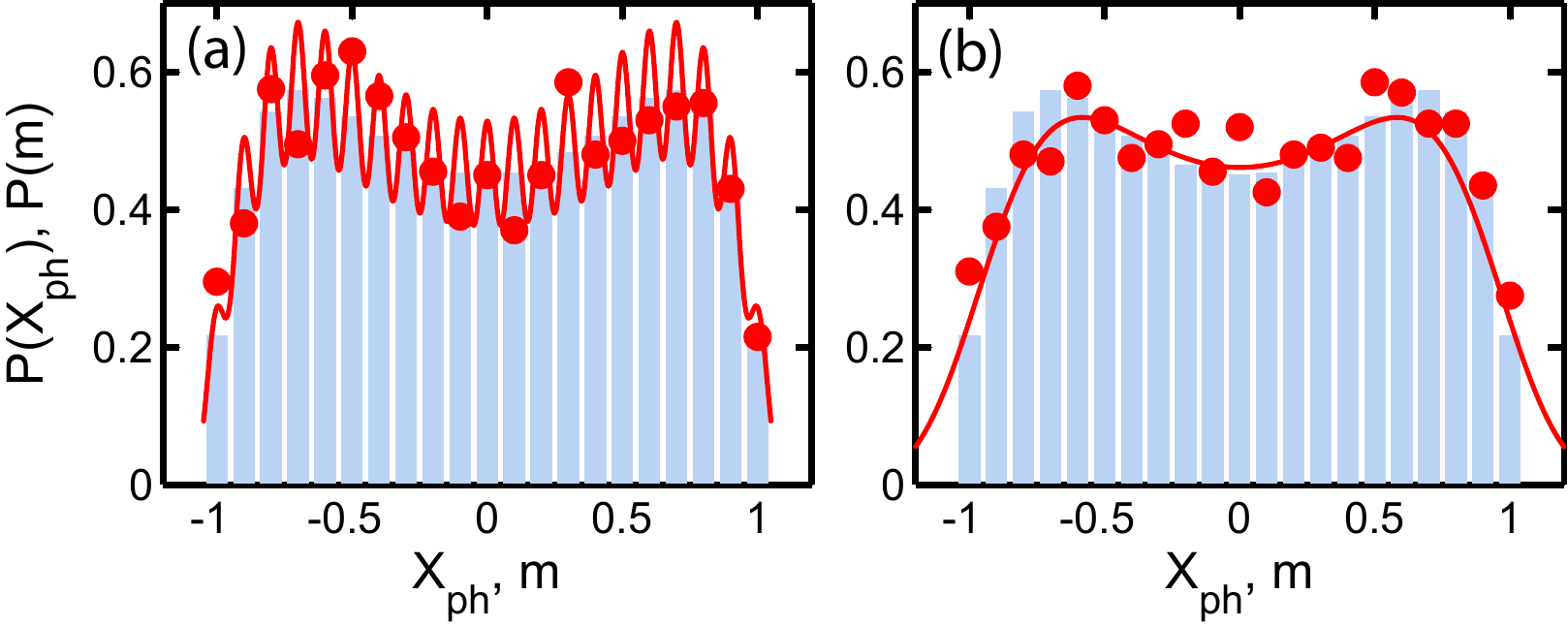}
\caption{(Color online) Comparison between rescaled light distribution (red solid line) and PDF of the order paramater (blue bar) in the critical phase in the Transverse Ising model for (a) $\tilde\kappa=1.0$, $\sigma_{ph}^2=1/2$ and (b) $\tilde \kappa=0.15$, $\sigma_{ph}^2=1/4$. The red circles correspond to the histogram obtained with a random variable following the light distribution for a number of shots $N_{shots}=2000$.} 
\label{Figure5}
\end{figure}

To evaluate the effectiveness of the method we compare the actual atomic spin distribution with the one of the output light. The  distance between both distributions decreases exponentially with $\tilde \kappa / \sigma_{ph}$. Thus, one could in principle improve the fidelity by increasing $\tilde \kappa$ or using squeezed light \cite{Mitchell2}. We show in Fig.~\ref{Figure5} the result for the transverse field Ising model for an optimal case $\tilde\kappa=1$ and $\sigma^2_{ph}=1/2$ and for a more realistic value of $\tilde\kappa=0.15$, but squeezed input light with $\sigma^2_{ph}=1/4$, recently achieved \cite{Mitchell2}. For the former, the light distribution faithfully follows the magnetization, whereas for the later, it only agrees qualitatively, but still captures the peaks. Experimentally, one would need to repeat the measurement a number of shots $N_{shots}\sim L^2$ to estimate the PDF. 

\section{Summary}
In conclusion, we have shown that the distribution of collective variables in spin models reveals relevant information of quantum phase transitions. We have shown that for a range of quantum phase transitions a non-Gaussian distribution of the order parameter is a clear signature of criticality, and that the scaling hypothesis holds. Finally we have proposed an experimental method for its measurement using light-matter interfaces, and discussed its  feasibility  for realistic values.\\

The authors want to thank D. Dagnino, A. Ferraro, G. Mussardo, A. Sanpera and R. Sewell for very useful comments. This work is supported by the UK EPSRC (EP/L005026/1), John Templeton Foundation (grant ID 43467), the EU Collaborative Project TherMiQ (Grant Agreement 618074). 

\section*{References}
\bibliographystyle{iopart-num}
\bibliography{biblio}
\end{document}